*Article*

# On-demand 5G Private Networks using a Mobile Cell

**André Coelho \*, José Ruela, Gonçalo Queirós, Ricardo Trancoso, Paulo Furtado Correia, Filipe Ribeiro, Helder Fontes, Rui Campos, Manuel Ricardo**

INESC TEC, Faculdade de Engenharia, Universidade do Porto, Portugal;
\* Correspondence: andre.f.coelho@inesctec.pt;

**Abstract:** This paper proposes the Mobile Cell (MC) concept for on-demand 5G private networks. The MC is designed to extend, restore, and reinforce 5G wireless coverage and network capacity on-demand, especially in areas with temporary communications needs or where it is costly or not possible to deploy a permanent fixed infrastructure. The design of the MC as well as the development, integration, and deployment in 5G private networks are discussed.

The Mobile Cell concept can be applied in multiple real-world environments, including seaports and application scenarios. Similarly to critical hubs in the global supply chain, seaports require reliable, high-performance wireless communications to increase efficiency and manage dynamic operations in real-time. Current communications solutions in seaports typically rely on Wi-Fi and wired-based technologies. Wired-based technologies lack the necessary flexibility for dynamic environments. Wi-Fi is susceptible to interference from other systems operating in the same frequency bands. An MC operating in a licensed, interference-free spectrum is a promising solution to overcome these limitations and provide improved Quality of Service when using the 5G technology.

## 1. Introduction

International seaports play a key role in trade, logistics, and the global supply chain. To improve operational efficiency and minimize costs, there is an increased interest in integrating novel technological solutions in seaports, including video surveillance, autonomous vehicles, and mission-critical communications. This will allow for environment-aware, data-driven aided decisions and adaptive management of the seaport operations in real-time [1–3].

When it comes to communications, many seaports, including the Port of Sines in Portugal, currently rely on Wi-Fi and wired-based technologies [4]. Wired-based technologies offer limited flexibility for dynamic networking scenarios. Wi-Fi is prone to interference from other systems operating in the same frequency bands, including interference from ships' radars when they dock at the seaport, leading to communications disruptions. This reduces network reliability and availability, compromising the overall performance.

The 5G technology, known for supporting enhanced Mobile Broadband (eMBB), massive Machine-Type Communications (mMTC), and Ultra-Reliable Low-Latency Communications (URLLC), operating in a licensed, interference-free spectrum, is a promising solution to meet heterogeneous requirements associated with key use cases [5]. Still, Radio Access Networks (RANs) typically rely on fixed Base Stations. In seaport scenarios, unexpected obstacles and machinery frequently compromise signal propagation. On the other hand, the dynamism of a multitude of operations carried out simultaneously imposes fluctuations in traffic demand.

On-demand wireless networks, with dynamic reconfiguration and repositioning capabilities, have emerged as a suitable solution to extend, restore, and reinforce wireless coverage where a fixed infrastructure is non-existent, insufficient or damaged. An on-demand network infrastructure leveraging 5G technology can improve the Quality of Service (QoS) offered to end-users in dynamic environments, such as seaports.

This paper describes the architecture, design, and development of a Mobile Cell (MC) for on-demand 5G private networks. The 5G MC enables the RAN to be reconfigured and repositioned over time, allowing for the extension of fixed communications infrastructures. This solution holds the potential to offer wireless connectivity anywhere in a seaport environment, especially in areas with temporary communications needs or where fixed infrastructures may not be feasible and cost-effective.

In this paper, the concept of MC refers to a mobile base station or a mobile base station relay, as described hereafter. The main goal of a 5G MC is to extend the coverage of a 5G network to zones where a User Equipment (UE) is out of reach of the fixed 5G infrastructure [6], which may be either a Public Land Mobile Network (PLMN) or a Non-Public Network (NPN).

As a downgraded version, an MC can be deployed as a nomadic node, placing it in a predefined, quasi-static position where it stays while needed, in order to reinforce capacity or provide 5G access where it is unavailable. This may be the case of temporary events and emergencies. In a seaport environment, an MC can be carried, for example, on ground vehicles, towboats, or drones, requiring a 5G wireless connection to the fixed infrastructure, as illustrated in Figure 1.

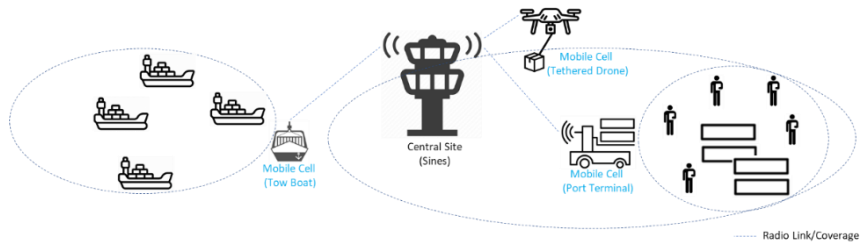

**Figure 1.** Mobile Cell (MC) carried by ground vehicles, towboats, or drones in a seaport scenario.

The MC aims at extending a 5G RAN [7] provided by a PLMN or an NPN to areas with limited wireless connectivity and network capacity, including offshore locations in seaport environments.

## 2. Reference Architecture

For a better understanding of the MC functional elements, the 5G RAN architecture is first described. The basic 5G RAN node is a base station called gNodeB (gNB). Besides the Radio Frequency (RF) unit, the 5G RAN functions implemented by gNBs are divided into various protocol layers, as shown in Figure 2: Physical (PHY), Medium Access Control (MAC), Radio Link Control (RLC), Packet Data Convergence Protocol (PDCP), and Radio Resource Control (RRC).

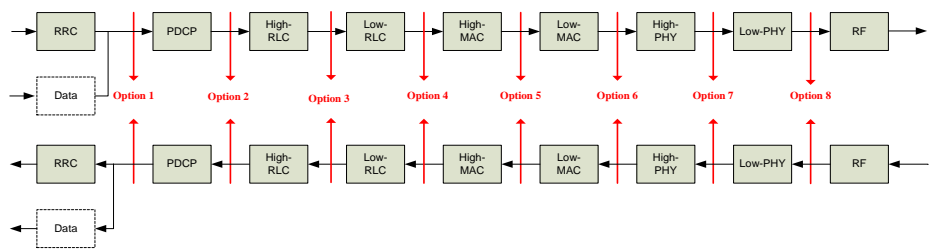

**Figure 2.** Function Split between central and distributed unit [8].

Figure 2 also shows eight possible functional split points that may be used to group functions into three functional units: a Central Unit (CU), a Distributed Unit (DU), and a Radio Unit (RU). A High-Level Split (HLS) separates the CU from the DU, while a Low-level Split (LLS) separates the DU from the RU. This allows multiple network configurations, with different physical and logical grouping and location of functions.

The 3rd Generation Partnership Project (3GPP) defined a Next Generation RAN (NG-RAN) architecture [9] composed of multiple gNBs. A gNB consists of one or more DUs controlled by a single CU, as illustrated in Figure 3.

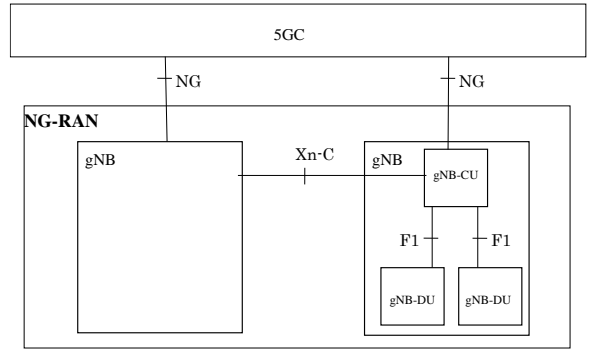

**Figure 3.** Overall NG-RAN Architecture [9].

For the CU/DU HLS, 3GPP adopted option 2, that is, the CU implements the RRC and PDCP protocol layers, and specified the IP-based F1 interface [10] for the communications between the CU and the DUs. Since this architecture does not impose any LLS, the RU functions are not explicitly represented.

The enhanced Common Public Radio Interface (eCPRI) [11], based on split 7.2, is a standardized LLS interface.

For the design of the MC, latency and throughput requirements on HLS and LLS interfaces were taken into account.

The MC is a mobile node hosted by a 5G network (PLMN or NPN), which provides the 5GC functions. It is either a full gNB or a relay node with DU and RU functions; in the latter case, the CU is part of the fixed 5G infrastructure. A 5G radio link is required to connect the MC to the hosting network.

### 3. Design and development

To achieve a proof of concept of the MC, we propose two alternative architectures: the MC is a full mobile gNB (no CU/DU split) or a mobile DU/RU relay controlled by a fixed CU (split 2).

In both solutions, it is necessary to establish IP connectivity between the MC and its home fixed network infrastructure, which includes 5G Core functions. For this purpose, a 5G Protocol Data Unit (PDU) session must be set-up by an overlay network (ON) between the MC and a User Plane Function (UPF) on the ON. The ON may be the home network (HN), for example a Standalone NPN, or a Mobile Network Operator (MNO) PLMN, when the MC is not within the range of a fixed gNB of the HN.

To set-up the PDU session, the MC acts as a regular UE, which is represented by a Mobile Termination (MT) function. First, a 5G wireless channel is set-up between the MT and a gNB on the ON and, once established, the MT PDU session is used to transparently carry all UE data and control traffic to the HN, through the MC and the ON.

The two MC-proposed solutions are agnostic to the architecture of the ON gNB(s), since the MC only requires the establishment of an MT PDU session. They differ on the endpoints (on the mobile platform and the fixed infrastructure) of the MT PDU Session.

An MC based on a no split architecture is a full gNB consisting of RU, DU, and CU entities placed on-board a mobile platform, while the 5GC functions reside on the fixed HN infrastructure. An MT function, acting as a UE, is employed to establish IP connectivity between the CU and the 5GC network through an ON, as explained before. This solution is depicted in Figure 4.

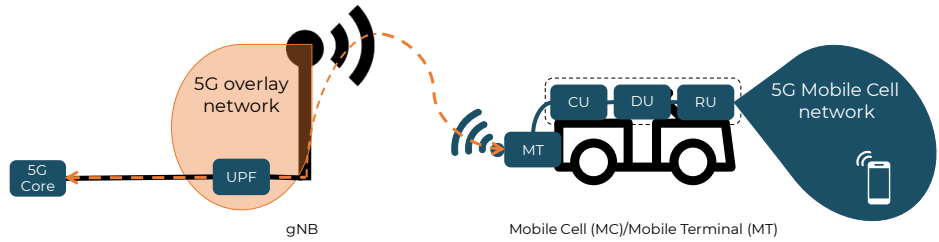

**Figure 4.** Solution based on CU, DU and RU on-board a mobile platform.

An MC based on the CU/DU split 2 architecture consists of a DU and an RU placed on-board a mobile platform, while the CU and the 5GC functions are part of the fixed HN infrastructure. Communications between the DU and CU entities take place over the F1 interface. First, the MT function must trigger the establishment of connectivity at IP level through an ON, as explained. This solution is depicted in Figure 5.

The connection between the CU and the 5GC network function is typically ensured by a wired link, such as optical fibre.

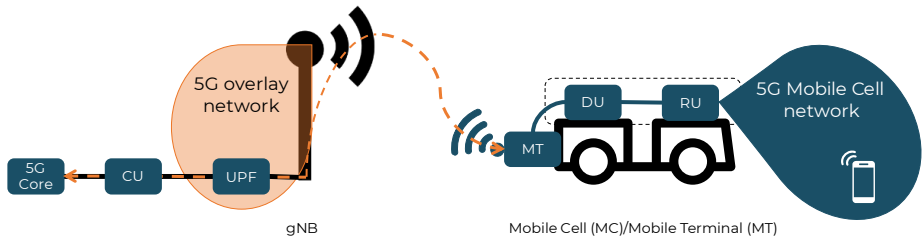

**Figure 5.** Solution based on DU and RU on-board a mobile platform and CU on the fixed infrastructure, close to the 5GC network.

## 4. Discussion

The MC solutions may be assessed and compared using as reference the architectures currently proposed by 3GPP.

The high-level specification of a mobile base station relay (MBSR) that takes advantage of the Integrated Access and Backhaul (IAB) architecture [9][12] is described in [8]. An IAB network is organised as a logical tree of IAB-nodes, with an IAB-donor at the root. An IAB node is composed of an IAB-DU and an IAB-MT function. The IAB-donor is based on the gNB CU/DU split 2 architecture and consists of a single IAB-donor-CU and one or more IAB-donor-DUs. The communications between the IAB-donor-CU and all DU entities take place over the F1 interface. The branches of the tree are wireless backhaul links (between a DU on a parent node and an MT on a child node), which are used to relay traffic between IAB-nodes and the IAB-donor. The DU entities also provide wireless access links to UE. An IAB MBSR is just a mobile IAB-node directly attached to the IAB-donor-DU, using the respective access link.

Unlike the MC split 2 architecture, the MBSR-DU is controlled by the same entity (IAB-donor-CU) that provides access to the 5G Core. Moreover, a Backhaul Adaptation Protocol (BAP) layer is required on all IAB DU and MT entities, on top of RLC/MAC/PHY layers; it maps IP packets (carried over the F1 interface) into RLC channels for relaying across backhaul links. For this reason, to directly attach to an IAB-donor, the MC split 2 architecture would have to be upgraded by adding BAP to the MC-MT stack.

Anyway, the current MC solutions allow attaching to an IAB network in the overlay mode – the IAB-donor is used as a conventional gNB, as also seen by UEs.

3GPP is currently promoting the study of a mobile gNB, under the designation of Wireless Access Backhaul (WAB) [13]. In architectural terms, this corresponds to the MC architecture with no CU/DU split. This solution has some advantages over the MC split 2 architecture: 1) it avoids the complexity of the F1 interface; 2) does not require an MC-CU on the fixed infrastructure; and 3) does not need tight control of the delay between the MC-DU and MC-CU as when connected by an ON. In addition, a 5G Core with basic functions may be deployed on the MC, thus providing some local services to UEs. This MC platform will be a valuable tool for research aligned with the WAB proposed objectives.

## 5. Conclusions

Private communications networks based on 5G technology are being deployed with success in seaports worldwide. In such environments, on-demand networks offer improved flexibility due to their reconfiguration and repositioning capabilities. In this context, mobile communications platforms based on the MC concept may play a new and important role, since they can be deployed on-demand to take advantage of their properties. They allow extending coverage and capacity of fixed communications infrastructures, potentially offering 5G wireless connectivity anytime and anywhere in the seaport.

The proposed MC has the potential to offer: 1) enhanced flexibility, allowing it to meet variable network requirements through dynamic repositioning; 2) scalability, as the number of MCs can be increased to extend network coverage without significant changes to the network infrastructure; and 3) cost-effectiveness, since MCs can reduce the need for fixed, permanent communications infrastructures with scarce use in some locations. The usage of MCs is especially relevant in areas with temporary communications needs or where fixed infrastructures may not be feasible or cost-effective.

For future work, we aim at implementing an MC for a real-world proof of concept in a seaport environment with representative use cases.

**Author Contributions:** Conceptualization, A.C., J.R., P.F.C., F.R., H.F., R.C., M.R.; methodology, A.C., J.R., F.R., H.F., R.C., M.R.; software, G.Q., R.T.; validation, A.C., G.Q., R.T.; investigation, all authors; writing—original draft preparation, A.C., J.R.; writing—review and editing, A.C., J.R., H.F., R.C., M.R.; supervision, M.R.; funding acquisition, F.R., H.F., R.C., M.R. All authors have read and agreed to the published version of the manuscript.

**Funding:** This work is co-financed by Component 5 - Capitalization and Business Innovation, integrated in the Resilience Dimension of the Recovery and Resilience Plan within the scope of the Recovery and Resilience Mechanism (MRR) of the European Union (EU), framed in the Next Generation EU, for the period 2021 - 2026, within project NEXUS, with reference 53.

**Institutional Review Board Statement:** Not applicable.

**Informed Consent Statement:** Not applicable.

**Data Availability Statement:** Not applicable.

**Conflicts of Interest:** The authors declare no conflicts of interest. The funders had no role in the design of the study; in the collection, analyses, or interpretation of data; in the writing of the manuscript; or in the decision to publish the results.

## References

1. Szymanowska, B.B.; Kozłowski, A.; Dąbrowski, J.; Klimek, H. Seaport Innovation Trends: Global Insights. Mar Policy 2023, 152, 105585, doi: https://doi.org/10.1016/j.marpol.2023.105585.
2. Neugebauer, J.; Heilig, L.; Voß, S. Digital Twins in Seaports: Current and Future Applications. In Proceedings of the Computational Logistics; Daduna Joachim R. and Liedtke, G. and S.X. and V.S., Ed.; Springer Nature Switzerland: Cham, 2023; pp. 202–218.
3. Sadiq, M.; Su, C.-L.; Parise, G.; Sayler, K. A Comprehensive Review on Roadmap to Seaports: Methodologies, Area of Use, and Purposes. In Proceedings of the 2024 IEEE/IAS 60th Industrial and Commercial Power Systems Technical Conference (I&CPS), Las Vegas, Nevada, 19 May 2024.

4. Han, Y.; Wang, W.; Chen, N.; Zhong, Y.; Zhou, R.; Yan, H.; Wang, J.; Bai, Y. A 5G-Based VR Application for Efficient Port Management. World Electric Vehicle Journal 2022, 13, doi: https://doi.org/10.3390/wevj13060101.

5. Charpentier, V.; Slamnik-Kriještorac, N.; Landi, G.; Caenepeel, M.; Vasseur, O.; Marquez-Barja, J.M. Paving the Way towards Safer and More Efficient Maritime Industry with 5G and Beyond Edge Computing Systems. Computer Networks 2024, 250, 110499, doi: https://doi.org/10.1016/j.comnet.2024.110499.

6. 3GPP System Architecture for the 5G System (5GS) (Release 18) 2024, TS 23.501.

7. 3GPP NR; NR and NG-RAN Overall Description; Stage 2 (Release 18) 2024, TS 38.300.

8. 3GPP Study on New Radio Access Technology: Radio Access Architecture and Interfaces (Release 14) 2017, TR 38.801.

9. 3GPP NG-RAN; Architecture Description (Release 18) 2024, TS 38.401.

10. 3GPP NG-RAN; F1 General Aspects and Principles (Release 18) 2024, TS 38.470.

11. Ericsson, A.; Huawei, T.Co.L.; NEC, C.; Alcatel, L.; Nokia, N. Common Public Radio Interface: ECPRI Interface Specification 2019, V2.0.

12. Zhang, Y.; Kishk, M.A.; Alouini, M.-S. A Survey on Integrated Access and Backhaul Networks. Frontiers in Communications and Networks 2021, 2, doi: https://doi.org/10.3389/frcmn.2021.647284.

13. NTT DOCOMO; AT&T Revised SID on Study on Additional Topological Enhancements for NR. 3GPP TSG RAN RP-240319 2024.